\documentclass[journal,12pt,onecolumn,draftclsnofoot]{IEEEtran}

\ifCLASSINFOpdf
\usepackage[pdftex]{graphicx}
\else
\fi
\usepackage{float} 
\usepackage{amsmath, icomma}

\usepackage{amsmath}
\usepackage{amsfonts}

\usepackage{subcaption}

\usepackage{xcolor}
\usepackage{hyperref}
\usepackage{algorithmic}

\usepackage{amssymb}
\usepackage{comment}

\usepackage{array}

\usepackage{pgfplots}
\pgfplotsset{compat=1.15}
\usepackage{mathrsfs}
\usetikzlibrary{arrows}

	\DeclareMathOperator{\Gal}{Gal}
	
	\DeclareMathOperator{\Aut}{Aut}

	\def\vF{\mathbb{F}}

	\def\vP{\mathbb{P}}

	\def\cO{\mathcal{O}}

	\newtheorem{theorem}{Theorem}[section] 
	\newtheorem{proposition}[theorem]{Proposition}
	
	\newtheorem{corollary}[theorem]{Corollary}
	\newtheorem{construction}[theorem]{Construction}
	\newtheorem{lemma}[theorem]{Lemma}

	\newtheorem{definition}{Definition}[section] 

	\newtheorem{remark}{Remark}

	\title{Optimal locally recoverable codes with hierarchy from nested $F$-adic expansions}
	
	\author{Austin~Dukes, Giacomo~Micheli and Vincenzo Pallozzi~Lavorante 
		\thanks{G. Micheli is with the Department of Mathematics, University of South Florida, ZIP, Tampa, US (e-mail: gmicheli@usf.edu).}
		\thanks{Manuscript received ...; revised ...}}

\begin{document}
		\maketitle
		%
		%
		\begin{abstract}
			In this paper we construct new optimal hierarchical locally recoverable codes. Our construction is based on a combination of the ideas of \cite{ballentine2019codes,sasidharan2015codes} with an algebraic number theoretical approach that allows to give a finer tuning of the minimum distance of the intermediate code (allowing larger dimension of the final code), and to remove restrictions on the arithmetic properties of $q$ compared with the size of the locality sets in the hierarchy. In turn, we manage to obtain codes with a wide set of parameters both for the size  $q$ of the base field,  and for the hierarchy size, while keeping the optimality of the codes we construct.

		\end{abstract}
\section{Introduction}\label{sec:intro}
Various classes of locally recoverable codes have received great attention in recent times due to their applications to cloud and distributed storage systems \cite{barg2017locally,barg2015locally,bartoli2020locally,dukes2022optimal,gopalan2014explicit, kamath2014codes,liu2018new,silberstein2013optimal,tamo2014family,
tamo2016bounds}.

In this paper we produce new optimal hierarchical locally recoverable codes (HLRC). HLRCs are suitable solutions that address the problem of recovering lost information in a distributed storage  system, and they  have been widely studied in \cite{ballentine2019codes,freij2016locally,sasidharan2015codes,tamo2016optimal}.

\noindent HLRCs allow to recover certain patterns of erasures by gradually looking at more components depending on the number of erasures that occurred. One can then design codes that recover one erasure by looking at at most $b$ other  components; $\lambda$ erasures by looking at $a$ other components; and  $d-1$ erasures by looking at at most $k$ components, where $k$ is the dimension of the code. This is impactful from a practical perspective, as one can deal with the most likely scenario (one erasure) in the optimal way, with the less likely scenario ($\lambda$ erasures) in an acceptable way, and still be able to recover $d-1$ erasures by accessing a large  number  of nodes.
Tuning these parameters in an efficient way depends on the reliability of the servers and the required efficiency of the system in terms of node retrieval.
One of the features that one would desire from this kind of  code  is that $\lambda$ is not too large, as the second most likely scenario is the failure of  only  a few other nodes more than 1 (and not too many others, that can anyway be recovered using the minimum distance). We address this problem by writing a sharper Singleton bound for this regime of parameters and  then constructing codes that achieve the bound.
Let us now define the main objects we will be treating in this paper.

\subsection{Definitions}
In the rest of the paper we will consider the occurence of either one, $\lambda$, or $d-1$ erasures, as these arise most commonly from applications (instead of the more general setting where one allows $\lambda_1$, $\lambda_2$, or $d-1$ erasures).
Let $n,k,b$ be positive integers with $k \leq n$. A locally recoverable code (LRC) $C$ having parameters $[n,k,b]$ is an $\mathbb{F}_q$-subspace of $\mathbb{F}_q^n$ of dimension $k$ such that if one deletes one component of any $v \in C$, this can be recovered by accessing at most $b$ other components of $v$. If $d$ is the minimum distance of the code, we will write that $C$ is an $[n,k,d,b]$ LRC.

We now give the following definition which will be useful in the rest of the paper.

\begin{definition}
Let $n$ be a positive integer, $C\subseteq \vF_q^n$ be a linear code, and $S$ be a subset of the set of indices $\{1,...,n\}$. We say that $C$ can tolerate $x$ erasures on $S$ if, whenever there are $x$ erasures on components of a codeword with indices belonging to $S$, the missing components can be recovered by looking at $|S|-x$ other coordinates in $S$.
\end{definition}

\begin{color}{black} In this paper we construct new locally recoverable codes with hierarchy of locality sets. Our Definition \ref{def:hierarchical} is equivalent to the one of hierarchical codes in \cite{ballentine2019codes} but we find it slightly easier to employ ours for practical situations, as we keep direct track of the size of the ``hierarchy''. \end{color}
\begin{definition}\label{def:hierarchical}
Let $n,k,d,b,a,\lambda$ be positive integers with $n>k$ \begin{color}{black} and $2\leq \lambda \leq b$ \end{color}. An $[n,k,d,b,a,\lambda]$ hierarchical locally recoverable code (HLRC) is an $[n,k,d]$-linear code such that
\begin{itemize}
\item $(a+\lambda) \mid n$,
\item $(b+1) \mid (a+\lambda)$,
\item the codeword indices are partitioned into $\ell$ \begin{color}{black}$\geq 1$\end{color} distinct sets $A_i$, each of size $a+\lambda$, such that $C$ tolerates $\lambda$ erasures on $A_i$ for every $i\in \{1,\dots,\ell\}$, and
\item each $A_i$ can be partitioned into $B_{i,j}$, each of size $b+1$, such that $C$ tolerates $1$ erasure on each $B_{i,j}$ for every $i\in \{1,\dots,\ell\}$ and every $j\in \{1,\dots, (a+\lambda)/(b+1)\}$.
\end{itemize}
\end{definition}

\subsection{Motivation}
Let us now briefly explain the motivation behind codes with hierarchical locality. Let $T$ be the time needed to replace a failed node. Suppose that a second node fails in the same locality set as the first node during the time $T$. An $[n,k,d,b]$ locally recoverable code will still need to access $k$ information symbols, as the $1$-locality procedure is not guaranteed to work anymore. However, an $[n,k,d,b,a,\lambda]$ hierarchical locally recoverable code only requires accessing at most $a$ information symbols. Since the failure of only a few nodes, say $\lambda < d-1$, is significantly more likely than the failure of $d-1$ nodes in the span of time $T$, it is convenient to have a code which addresses separately the case in which only $\lambda$ nodes fail.
The codes in \cite{ballentine2019codes} address this issue, but they are restricted to certain $\lambda$'s, as we explain in subsection \ref{sec:comparison} and moreover in many cases they require restrictions on the arithmetic of $q$ and the size of the hierarchy (see for example the case of power functions in \cite[Section IV.A, Example]{ballentine2019codes}).

\subsection{Our contribution}

In this paper we provide new constructions of optimal codes with hierarchical locality and an improved bound for HLRC for a special set of parameters. Our construction is based on the ideas in \cite{ballentine2019codes} combined with powerful techniques from algebraic number theory, allowing us to remove arithmetic restrictions on the size of the hierarchy compared with $q$ or $q-1$.

\emph{Structure of the paper:} 
\begin{itemize}
\item In Section \ref{sec:intro} and its subsections we explain the basic coding theoretical definitions and provide the practical motivations for the study of such codes.
\item In Section \ref{sec:singleton}, for some regime of parameters, we provide a stronger Singleton bound than the one already present in the literature for HLRC \cite{sasidharan2015codes}. Our bound beats the previous bound for an infinite set of parameters (see for example Remark \ref{rem:boundbeat}).
\item In Section \ref{sec:codereal} we achieve our new bound with a new construction of HLRC that covers a set of parameters that are not available using previous constructions (see subsection \ref{sec:comparison}). 
In subsection \ref{sec:toyexample} we construct one of our codes to show how a generator matrix looks like in practice. 
\item In Section \ref{sec:existresults} we show that our codes are constructible without requiring arithmetic arithmetic restriction of $q,q-1$, the locality parameters, and the sizes of the sets in the hierarchy.
\item Using the existential results provided in Section \ref{sec:existresults}, in Section \ref{sec:practical} we provide some practical choice of parameters for codes with large  length.

\end{itemize}

\section{An improved bound for hierarchical locally recoverable codes}\label{sec:singleton}

\subsection{The Singleton bound for $[n,k,d,b,a,\lambda]$ hierarchical LRCs with $\lambda\leq b$. }\label{sec:singleton2}
		Let $ \textit{M}_{m \times n}(q)$ denote the set of all matrices of dimension $m \times n$ defined over $\mathbb{F}_q$. The following is a well-known proposition, but	we include a proof for completeness.
		\begin{proposition}
			\label{fact:optim}
			Let $C$ be an $ [n,k,d]_q $ code with generator matrix $\mathcal{G} \in \textit{M}_{k \times n}(q)$ and let  $S \in \textit{M}_{k \times t}(q)$ be a submatrix of $\mathcal{G}$. If $rk(S)\leq k-1$ then $t \leq n-d$.
		\end{proposition}
		\begin{IEEEproof}
			Let $ S=[S_1, \dots, S_t] $, where $S_i$ is a column of $\mathcal{G}$ for $i\in\{1,\dots,t\}$. Define $\tilde{S}\colon \mathbb{F}_q^k \to \mathbb{F}_q^t$ such that $x \mapsto \tilde{S}(x)=xS=[xS_1,\dots, xS_t]$. Since $rk(S) \leq k-1$ and we can write $\tilde{S}(x)=x_1 R_1+\dots x_k R_k$, where $R_i$ are the rows of $S$, there exists $x' \in \mathbb{F}_q^k$ such that $\tilde{S}(x')=0$. Assuming without loss of generality that $S$ consists of the first $t$ columns of $\mathcal{G}$, there exists a codeword $c=[\tilde{S}(x'),y_{t+1},\dots,y_n]$ whose weight equals $n-t$. Hence $d \leq n-t$.
		\end{IEEEproof}

To help the reader understand the more complex bound we propose on hierarchical LRCs, we include here a proof of the standard Singleton bound for LRCs, which does not make formal use of an algorithm like in other proofs in the literature (the ideas in the proof are the same).
		\begin{corollary}\label{cr:loc_r}
			Let $C$ be a $[n,k,d,b]$ LRC. Then \[ k-1+\left\lfloor\frac{k-1}{b}\right\rfloor  \leq n-d.\]
		\end{corollary}
		\begin{IEEEproof}
			We know that every set of $b+1$ columns has rank $b$ by the locality condition. This means that we can choose a set $S$ of $ \left\lfloor \frac{k-1}{b} \right\rfloor(b+1) + \left\{\frac{k-1}{b}\right\}b$ columns in such a way that $rk(S) \leq k-1$ (here $\{x\}=x-\lfloor x\rfloor$).
			Thus, by applying Proposition \ref{fact:optim}, we have the following:
\begin{equation*}
\left(\left\lfloor \frac{k-1}{b}\right\rfloor+\left\{\frac{k-1}{b}\right\} \right)b+\left\lfloor\frac{k-1}{b}\right\rfloor = k-1+\left\lfloor\frac{k-1}{b}\right\rfloor  \leq n-d.
\end{equation*}
\end{IEEEproof}
Notice that the above is equivalent to the well-known bound $d\leq n-k-\lceil k/b \rceil+2$.

We aim to generalize the bound in Corollary \ref{cr:loc_r} when $C$ is an $[n,k,d,b,a,\lambda]$ HLRC. The key observation is that one can partition the columns of the generator matrix into $\ell$ sets of $a+\lambda$ columns so that each set has rank strictly less than $a$, and each set of $a+\lambda$ columns can be partitioned further into sets of $b+1$ columns so that each set has rank at most $b$. 
To see this, note that each set $S_i$ of $a+\lambda$ columns (corresponding to $A_i$) can be divided into $\beta =(a+\lambda)/(b+1)$ sets, say $S_{i,j}$ for $j \in \{1,\dots,\beta\}$, of $b+1$ columns (corresponding to $B_{i,j}$) with rank at most $b$ for $i\in \{1,\dots,\ell\}$ by the definition of the code.
		Now, in the first set $S_{i,1}$ we have $\lambda$ columns which are in the span of the other $a$ columns in $S_i$. This means that we can choose $b+1-\lambda$ columns from $S_{i,1}$ and $b$ columns from each of the other $S_{i,j}$, with $j \ne 1$, and be able to recover any $\lambda$ of the $a+\lambda$ columns in $S_i$. Therefore, the rank of each $S_i$ is at most \[\rho:=\big[\underbrace{(a+\lambda)/(b+1)-1}_{\beta -1}\big]b+(b+1-\lambda)\leq a.\]
		
\begin{theorem}\label{th:optimality}
Let $C$ be a $[n,k,d,b,a,\lambda]$ HLRC with $\lambda\leq b$, and let $\rho=b(a+\lambda)/(b+1)-(\lambda-1)$. Then
\begin{equation}\label{eq:UpBound}
\Big\lfloor \frac{k-1}{\rho} \Big\rfloor (a+\lambda) + k_1+ \Big\lfloor\frac{k_1}{b}\Big\rfloor \leq n-d,
\end{equation}
where $k-1 \equiv k_1 \pmod{\rho}$ and $0 \leq k_1 < \rho$.
\end{theorem}
\begin{IEEEproof}
Given $\Big\lfloor \frac{k-1}{\rho} \Big\rfloor$ locality sets $A_i$, say $i \in \{ 1, \dots, \Big\lfloor \frac{k-1}{\rho} \Big\rfloor\}$, denote by $S$ the set of the corresponding columns of the generator matrix $\mathcal{G}$ of the code. Thus, $|S|=\Big\lfloor \frac{k-1}{\rho} \Big\rfloor (a+\lambda)$ and by the above discussion, we have $rk(S)=\rho \Big\lfloor \frac{k-1}{\rho} \Big\rfloor  \leq k-1$.
This allows us to add more columns to $S$ until the rank equals $k-1$ using a smaller locality set. More precisely, we can always choose a set of the remaining columns of $\mathcal{G}$, say $S_1$, of size $\lfloor\frac{k_1}{b}\rfloor(b+1) + \{\frac{k_1}{b}\}b$, such that $rk(S_1) = k_1$ (explicitly, $S_1$ is the union of columns which correspond to $B_{i,j}$, for $j \in \{1,\dots,\lfloor\frac{k_1}{b}\rfloor \}$) . Hence, \[rk(S\cup S_1) = \Big\lfloor \frac{k-1}{\rho} \Big\rfloor \rho +k_1=k-1,\]
by the definition of $k_1$.
Applying Proposition \ref{fact:optim} we have
\[\Big\lfloor \frac{k-1}{\rho} \Big\rfloor (a+\lambda)+\Big\lfloor\frac{k_1}{b}\Big\rfloor(b+1) + \biggr\{\frac{k_1}{b}\biggr\}b \leq n-d.\]
Now, since $\frac{k_1}{b}=\lfloor\frac{k_1}{b}\rfloor + \{\frac{k_1}{b}\}$ we have
\[\Big\lfloor \frac{k-1}{\rho} \Big\rfloor (a+\lambda) + k_1+ \Big\lfloor\frac{k_1}{b}\Big\rfloor \leq n-d.\]
\end{IEEEproof}

		\begin{definition}\label{def:optimal}
			We say that an $[n,k,d,b,a,\lambda]$ HLRC is \textit{optimal} if its minimum distance attains the upper bound in \eqref{eq:UpBound}, i.e., if 
			\[d=n-\left(\Big\lfloor \frac{k-1}{\rho} \Big\rfloor (a+\lambda) +k_1+ \Big\lfloor\frac{k_1}{b}\Big\rfloor\right),\]
			for $k-1 \equiv k_1 \pmod{\rho}$ and $0 \leq k_1 < \rho$.
		\end{definition}

\begin{remark}\label{rem:boundbeat}
Notice that our bound improves upon the bound in \cite{sasidharan2015codes} for infinitely many parameters, but ours holds only for $\lambda\leq b$. In fact, for any length $n$, and for parameters $k=6$, $a=4$, $r_1=\rho=3$, $r_2=b=2$, $\delta_1=\lambda+1=3$ and $\delta_2=2$ \cite[Theorem 2.1]{sasidharan2015codes} gives 
 $d\leq n-8$ when instead our bound gives $d\leq n-9$. The moral reasons for this are that we are taking into account a finer arithmetic of the parameters which involves the reduction of the dimension modulo the upper level hierarchical locality, and we are restricting to the case in which the number of nodes that we simultaneously erase is strictly smaller than the size of \begin{color}{black} the smaller locality set \end{color}.
\end{remark}

		\section{Our construction of optimal HLRCs using nested $f$-adic expansions}\label{sec:codereal}
		
		\subsection{Main tool for the construction}

		\begin{lemma}\label{lm:f(h(X))_split}
			Let $f,h \in \mathbb{F}_q[X]$ be non-constant polynomials. Suppose there is some $t_0 \in \mathbb{F}_q$ such that $f(h(X)) - t_0$ splits completely (i.e., factors into $(\deg f) (\deg h)$ distinct factors) over $\mathbb{F}_q$. Then the set of roots of $f(h(X)) - t_0$, say $A_0$, can be partitioned into sets $B_1, \ldots, B_{\deg f} \subseteq \mathbb{F}_q$ which satisfy the following:
			\begin{itemize}
				\item[$\bullet$] $h(B_i) = c_i \in \mathbb{F}_q$ for each $1 \leq i \leq \deg f$,
				\item[$\bullet$] the cardinality of each $B_i$ is $\deg h$, and
				\item[$\bullet$] $h(B_i) \neq h(B_j)$ whenever $i \neq j$.
			\end{itemize}
		\end{lemma}
		\begin{IEEEproof}
			By the hypothesis we may write $\displaystyle f(h(X))-t_0 = \prod_{i=1}^{(\deg f) (\deg h)} (X-x_i)$ for distinct elements $x_1, \ldots, x_{(\deg f) (\deg h)} \in \mathbb{F}_q$. Notice now that if $f(h(X)) - t_0$ splits completely, then $f(X)-t_0$ splits completely.  If we let $\alpha_1, \ldots, \alpha_{\deg f} \in \overline{\mathbb{F}}_q$ be the (distinct) roots of $f(X)-t_0$, then we may also write $\displaystyle f(h(X))-t_0 = \prod_{i=1}^{\deg f} (h(X) - \alpha_i) \in \overline{\mathbb{F}}_q[X]$. Combining these two factorizations and relabeling the $x_i$ appropriately yields $$\prod_{i=1}^{\deg f} \prod_{j=1}^{\deg h} (X-x_{i,j}) = \prod_{i=1}^{\deg f} (h(X)-\alpha_i),$$ where $\displaystyle \prod_{j=1}^{\deg h} (X-x_{i,j}) = h(X) - \alpha_i$ for each $1 \leq i \leq \deg f$. In particular, it follows that $\alpha_i \in \mathbb{F}_q$ for each $i$. Write $B_i = \{x_{i,j} : 1 \leq j \leq \deg h\}$. Then we have $h(B_i) = \alpha_i$ for each $i$, proving the first statement. The second and third statements both follow from the fact that the $x_{i,j}$'s are pairwise distinct and the $B_i$'s are respectively  and disjoint.
		\end{IEEEproof}

		\begin{definition}
			For $f,h \in \mathbb{F}_q[X]$, we say that a set $A \subset \vF_q$ is a \textit{nest} for $(f,h)$ if $A$ is the set of preimages of $t_0 \in \vF_q$ such that  $f(h(X))-t_0$ is totally split.
			
			\noindent Furthermore, we say that $B \subset A$ is a \textit{sub-nest} if $h$ is constant on $B$ and $|B|=\deg h$.
		\end{definition}

		\subsection{The main construction}
		We present a general method of constructing linear codes with the nested locality property. Later we will show that these codes are optimal in the sense of Section III. In line with the notion of $(r,\ell)$-good polynomials in \cite{micheliIEEE}, we now begin defining our \textit{nested} polynomials.
		\begin{definition}[\emph{$\ell$-nested}]
			Let $f,h \in \vF_q[X]$ and let $\ell$ be a positive integer. Then $f$ and $h$ are said to be $\ell$-nested if $f(h(X))-t_0$ splits completely over $\vF_q$ for at least $\ell$ elements $t_0 \in \vF_q$. 
		\end{definition}
		
		\begin{remark}\label{rem:subnest}
			Note that if $f$ and $h$ are $\ell$-nested, then from Lemma \ref{lm:f(h(X))_split} there exist $A_1,\dots,A_\ell$ distinct nests for $(f,h)$ such that
			\begin{itemize}
				\item for any $i \in \{1,\dots,\ell\}$, $f(h(A_i))=\{t_i\}$ for some $t_i \in \vF_q$,
				\item $ |A_i| =\deg f \deg h$,
				\item $A_i \cap A_j = \emptyset$ for any $i \ne j$, and
				\item each $A_i$ can be partitioned into sub-nests $B_{i,j}$ for $(f,h)$.
			\end{itemize}
			Those properties will be the key of our next construction.
		\end{remark}
		
		\begin{construction}[\emph{HLRC}]\label{con:LRC}
			Let $f,h \in \vF_q[X]$ be $\ell$-nested, with $3\leq\deg h= b+1$ and $\deg f(X) = \frac{a+\lambda}{b+1}$ for some integer $2\leq \lambda \leq b$, and let $\mathcal{A}=\cup_{i=1}^\ell A_i$, where $ \{A_1,\dots,A_\ell\} $ is a set of nests for $(f,h)$. 
			
			\noindent For a positive integer $s \geq 1$, consider the set $\mathcal{M}$ of polynomials of the form
			
			\begin{equation}\label{eq:message}
				m(X)=\sum_{i=0}^{s}\left[\left(\sum_{j=0}^{\deg f-2}g_{i,j}(X)h(X)^j\right)+\tilde{g}_{i}(X)h(X)^{\deg f -1} \right]f(h(X))^i,
			\end{equation}
			where $g_{i,j} \in \vF_q[X]_{\leq \deg h -2}$ and $\tilde{g}_{i} \in \vF_q[X]_{\leq\deg h-\lambda-1}$.

			\noindent Let $n=(\deg f \deg h) \ell$ and let $k$ be the dimension of $\mathcal{M}$ as an $\vF_q$-vector space. 
			
			\noindent Define
			\begin{equation}\label{eq:Code}
				\mathcal{C}:=\{(m(x), x \in \mathcal{A}) \mid m \in \mathcal{M} \}.
			\end{equation}
			We will prove that $\mathcal{C}$ is an optimal $[n,k,b,a,\lambda]$-HLRC over $\vF_q$.
		\end{construction}

		\subsection{Locality}\label{sec:locality}
		
		Since we evaluate at $n$ distinct points of $\mathbb{F}_q$, we need $q \geq n$. Write $n = (a + \lambda)(s+1)$ and recall that $b+1$ divides $a + \lambda$.
		
		Take $a \in \mathbb{F}_q^k$ and write Enc$_C(a) = c = c_{i,j_1,j_2}$ for $1 \leq i \leq s+1$, $1 \leq j_1 \leq (a + \lambda)/(b + 1)$, $1 \leq j_2 \leq b + 1$. Note that the index $i$ determines a nest $A_i$, $j_1$ determines a sub-nest $B_{i,j_1}$  and $j_2$ a precise element of the sub-nest considered, which is denoted by  $ c_{i,j_1,j_2}  $ indeed.  We begin by showing that the code $C$ described in Construction \ref{con:LRC} allows one to recover a single missing component of $c$ by accessing at most $b$ other components of $c$.
		
		Fix $\ell \geq 1$ and let $f,h \in \mathbb{F}_q[X]$ be the $\ell$-nested polynomials from which $C$ is obtained. Write $\mathcal{A} = \{A_1, \ldots, A_{\ell}\}$ with $A_i = \displaystyle \bigsqcup_{j=1}^{\deg f} B_{i,j_1}$ and $B_{i,j_1} = \{x_{i,j_1,j_2} : 1 \leq j_2 \leq b+1\}$ as in Remark \ref{rem:subnest}.
		
		Without loss of generality, assume that the missing component is $c_{1,1,b+1} = m_a(x_{1,1,b+1})$, where $m_a \in \mathcal{M}$. Observe immediately that because both of $f \circ h$ and $h$ are constant on $B_{1,1}$, the restriction $m_a |_{_{B_{1,1}}}$ can be written as a polynomial of degree $\max \{\deg h - 2, \deg h - \lambda-1\} = \deg h - 2 = b - 1$. Since $x_{1,1,j_2} \in B_{1,1}$ for each $j_2$, we have that $m_a |_{_{B_{1,1}}}(x_{1,1,j_2}) = m_a(x_{1,1,j_2}) = c_{1,1,j_2}$. Using Lagrange interpolation on the points $(x_{1,1,j_2}, c_{1,1,j_2})$ for $1 \leq j_2 \leq b$, we obtain a polynomial $\Delta^{B_{1,1}}$ of degree $b-1$ which agrees with $m_a |_{_{B_{1,1}}}$ at $b$ distinct points, so the two polynomials must be equal. Thus we can recover $c_{1,1,b+1}$ by evaluating $\Delta^{B_{1,1}}$ at the element $x_{1,1,b+1}$.
		
		Let us now consider the case of $\lambda$ erasures (in the practical example we will take $\lambda=2$, as that is the second most likely scenario of failures. Among these $\lambda$ erasures,  the erasures which are isolated in locality sets $B_{i,j}$ can be recovered by using the $1$-locality, so the interesting case is when multiple erasures occur in the same $B_{i,j}$. Let us assume that $\lambda \geq 2$ erasures occur in the same locality set $B_{i,j}$. In this case, since $f \circ h$ is constant on $A_i$, the restriction $m_a |_{_{A_i}}$ is a polynomial of degree $\deg f \deg h-\lambda-1=a+\lambda-\lambda-1=a-1$. Thus Lagrange interpolation on a set of $a$ points of $A_i$ on which no erasure occurred yields a polynomial $\Delta^{A_i}$ which agrees with $m_a$ on all of $A_i$. Hence the missing components can be obtained by evaluating $\Delta^{A_i}$ at each of the corresponding locations in $A_i$.

		\subsection{Optimality of the code}\label{subsec:optcode}
		We dedicate this subsection to proving the optimality of our code $\mathcal{C}$.
		Therefore, we will be computing the values of $k$ and $d$.
		\begin{lemma}\label{lm:k}
			Let $\mathcal{C}$ be the code in \eqref{eq:Code}. Then 
			\[k=(s+1) ((\deg f-1) (\deg h-1)+\deg h-\lambda)\]
		\end{lemma}
		\begin{IEEEproof}
			Since in particular $\deg (g_{i,j} h^j+\deg \tilde{g}_i h^{\deg f -1})\leq \deg(f \circ h)$ and $\deg g_{i,j} , \deg \tilde{g}_i \leq \deg h$, by uniqueness of $F$-adic expansion both for $F=f\circ h$ and for $F=h$, we have
			\[k=\dim_{\mathbb{F}_q} \mathcal{M}=(s+1)((\deg f-1)(\deg h -1)+(\deg h -\lambda)),\]
			as we wanted to prove.
		\end{IEEEproof}
		
		\begin{lemma}\label{lm:d}
			Let $\mathcal{C}$ be the code in \eqref{eq:Code}. Then $d\geq n - \delta$, for
				\[\delta= (s+1) \deg h \deg f - \lambda-1 \]
		\end{lemma}
		\begin{IEEEproof}
			A lower bound for the minimum distance is obtained by subtracting $\delta$ from $n$, where $\delta$ is the upper bound for the maximum number of zeros of $m \in \mathcal{M}$. 
			We compute
		\begin{equation}
			\begin{split}
				\delta=&((\deg f \deg h) s+(\deg f-1) \deg h+\deg h-\lambda-1) \\
				=&	(s+1) \deg h \deg f - \lambda-1, 
			\end{split}
		\end{equation}
			and this proves the claim.
		\end{IEEEproof}
		
		\begin{theorem}
			Let $\mathcal{C}$ be the code obtained by using Construction \ref{con:LRC}. Then $\mathcal{C}$ is an optimal $[n,k,b,a,\lambda]$ HLRC.
		\end{theorem}
	\begin{IEEEproof} Let $\rho =(a+\lambda)/(b+1)b-(\lambda-1)$ and $k_1= k-1-\Big\lfloor \frac{k-1}{\rho} \Big\rfloor \rho$. Moreover, we recall that  $a+\lambda=\deg f \deg h$ and $\deg h =b+1$. Let $d'$ denote the optimal distance, such that \[\delta'=n-d' =\left(\Big\lfloor \frac{k-1}{\rho} \Big\rfloor (a+\lambda) +k_1+ \Big\lfloor\frac{k_1}{b}\Big\rfloor\right).\]
		
		\noindent Note that \[\Big\lfloor \frac{k-1}{\rho} \Big\rfloor = \Big\lfloor\frac{1}{-\deg f \deg h+\deg f+\lambda-1}+s+1\Big\rfloor=s,\]
		since $\lambda \leq \deg h -1$, and 
		\[ \left\lfloor\frac{k_1}{b}\right\rfloor=\left\lfloor \deg f-\frac{\lambda}{\deg h-1}\right\rfloor=\deg f -\left \lceil\frac{\lambda}{\deg h-1}\right\rceil,\]
		in fact $k_1=\deg f (-(b+1) s+\deg h (s+1)-1)-\lambda=\deg f(\deg h-1)-\lambda$.
		By using the results of Lemma \ref{lm:k}, \ref{lm:d}, we have 
		
		\begin{equation}
			\begin{aligned}
		\delta-\delta'&= (s+1) \deg h \deg f - \lambda-1-	s \deg f \deg h -k_1- \left\lfloor\frac{k_1}{b}\right\rfloor \\
		&=\deg f \deg h-\lambda-1-(\deg f(\deg h-1)-\lambda)-\deg f +\left \lceil\frac{\lambda}{\deg h-1}\right\rceil\\
				&=\left\lceil\frac{\lambda}{\deg h-1} \right\rceil -1
			\end{aligned}
		\end{equation}
		and since $\Big\lceil\frac{\lambda}{\deg h-1} \Big\rceil -1=0$ for $\lambda \leq \deg h -1$, the code is optimal.
	\end{IEEEproof}

\subsection{Comparison with the optimal hierarchical RS-like code in \cite{ballentine2019codes}}\label{sec:comparison}

A construction of optimal hierarchical LRCs for a certain set of parameters is presented in \cite[Proposition IV.2]{ballentine2019codes}. Let us fix the parameters for which that construction exists, i.e., $r_1=sr_2$ (we note that we do not require such a constraint, but that even in this scenario we show that we can construct codes that are not available from \cite[Proposition IV.2]{ballentine2019codes}\begin{color}{red}) \end{color}. The set of parameters of the codes in \cite[Proposition IV.2]{ballentine2019codes}, given also in our notation, is as follows: 
\begin{itemize}
\item the length of the codes in both settings is $n$,
\item each small locality set (at the bottom level of the hierarchy)  has size $r_2+1$, so in our case each has size $b+1$,
\item their $\nu$ is our $a+\lambda$,
\item the middle code has distance $r_2 + 3$ and hence can tolerate $r_2 + 2$ erasures, so their $r_2 + 2$ corresponds to our $\lambda$,
\item their $r_1$ is our $\rho$,
\item the code is optimal, with distance \begin{color}{black} $d=n-t(r_1+r_2+1+s)+r_2+3$, for some $t,s$,
\item the two level hierarchy have locality parameters $(r_1,r_2+3)$ and $(r_2,2)$. \end{color}
\end{itemize}
This shows immediately that our class of codes is different from the codes in \cite[Proposition IV.2]{ballentine2019codes}. In fact, the optimality of our codes strongly relies on the assumption $\lambda \leq r_2$, which is not the case in the construction in \cite[Proposition IV.2]{ballentine2019codes}, in which instead $\lambda=r_2+2$.
It follows that our class of codes contains codes which are not covered by this construction, as we can construct optimal hierarchical codes with two level hierarchy having locality parameters  $(r_1,\lambda)$ and $(r_2,2)$, for any $\lambda \leq r_2$, such as $\lambda=2$. 

We emphasize that in \cite{sasidharan2015codes} it is necessary  to set a fixed $\lambda=r_2+2$ since in this way one can reach optimality using the bound in \cite[Theorem 2.1]{sasidharan2015codes}, while, using our improved bound and enhancing the construction in \cite{ballentine2019codes},  one is allowed more flexibility as we explained.
Moreover, we will see in detail how to construct our codes \begin{color}{black} without the arithmetic restrictions appearing in the examples which use monomials or linearized functions\end{color} (see Section \ref{sec:existresults}).

	For a better comparison and to simplify the understanding, in the next paragraph, we will still use monomials for the toy example, even if it is not a requirement as we explain in Section \ref{sec:existresults}.
		\subsection{Toy example}\label{sec:toyexample}
		
		Suppose one desires a code over $\vF_{19}$ of dimension $6$ which can recover $1$, $2$, and $8$ lost nodes by accessing at most $2$, $4$, and $6$ other nodes, respectively (i.e., the distance of the code equal $9$).
		This is not possible using standard Tamo-Barg construction since, to recover more than $1$ node, one would need to access as many nodes as the dimension of the code, that is, $6$ nodes. 
		Another option is to consider codes with availability using an orthogonal partition of the multiplicative group of $\vF_{19}$ that includes $C_3$ (as one wants the locality to be $3$). But this does not work in this case either as the only other option is $C_9$ and $C_3\subseteq C_9$ (since $\vF_{q}^{*}$ is cyclic for any prime power $q$). Moreover \cite[Proposition IV.2]{sasidharan2015codes} does not hold for $\lambda=2$.
		
		Our construction instead provides a code that allows these recovery cababilities and is information theoretically optimal in the sense of the Singleton bound in Subsection \ref{sec:singleton2}.

		Suppose we choose $f=X^2$ and $h=X^3$ (so $b=2$ and $a=4$). Then a general information polynomial is given by
		\[m(X)=\sum_{i=0}^{1}{\Big[g_{i}(X)+\tilde{g}_{i}(X)h(X)\Big]f(h(X))^i},\]
		for $g_{i} \in \vF_q[X]_{\leq 1}$ and $\tilde{g}_{i} \in \vF_q[X]_{\leq 0}$. In particular the $\tilde{g}_{i}(X)=\tilde{g}_{i}$ are constants (notice that the internal sum in $j$ in \eqref{eq:message} disappears as $\deg(f)=2$). Therefore, by evaluating the messages at the preimage of the the $3$ totally split places of $x^6=f\circ h$, we get a code of length $n=18$, dimension $k=6$, $b=2$ and $a=4$. Notice that this code can recover $1$ erasure by looking at $b=2$ other nodes. Moreover, if two erasures occur, we have two possibilities: either the erasures occur in the same nest for $(f,h)$, in which case one needs to access (in the worst case scenario) at most $4$ other nodes, or the erasures occur in different nests, in which case one can use twice the locality (that is, $2$) to recover each node so that one again needs to access at most $4$ other nodes. Since we are evaluating polynomials of degree at most $9$, the distance of the code is $18-9 = 9$ and therefore one has a fault tolerance of $8$ erasures.
		Practically, given those $18$ nodes, we are looking at the following disposition of hierarchy:

%

		\begin{figure}[H]
			\centering
			\includegraphics[width=\textwidth]{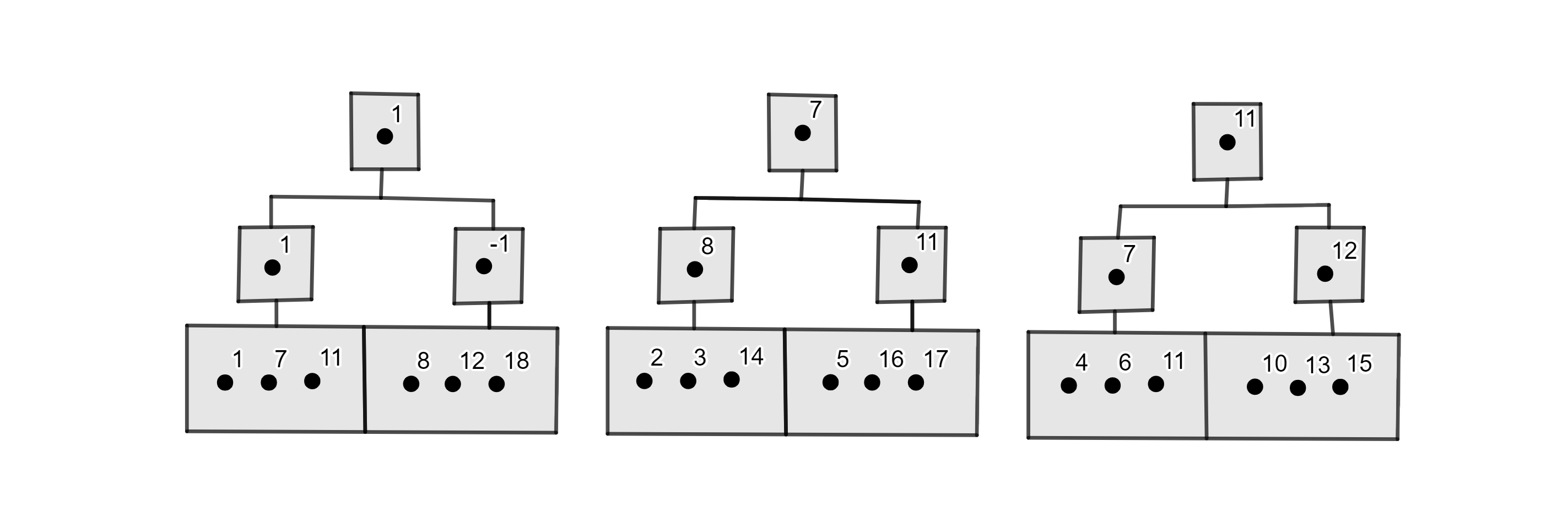}
			\caption{point-sets}
			\label{fig:float}
		\end{figure}

%
		
		which corresponds to the following matrix:
\setcounter{MaxMatrixCols}{20}
		\begin{equation*}
			\begin{pmatrix}
				1 & 1 & 1 & 1 & 1 & 1 & 1 & 1 & 1 & 1 & 1 & 1 & 1 & 1 & 1 & 1 & 1 & 1 \\
				1 & 7 & 11 & 8 & 12 & 18 & 2 & 3 & 14 & 5 & 16 & 17 & 4 & 6 & 9 & 10 & 13 & 15 \\
				1 & 1 & 1 & 18 & 18 & 18 & 8 & 8 & 8 & 11 & 11 & 11 & 7 & 7 & 7 & 12 & 12 & 12 \\
				1 & 1 & 1 & 1 & 1 & 1 & 7 & 7 & 7 & 7 & 7 & 7 & 11 & 11 & 11 & 11 & 11 & 11 \\
				1 & 7 & 11 & 8 & 12 & 18 & 14 & 2 & 3 & 16 & 17 & 5 & 6 & 9 & 4 & 15 & 10 & 13 \\
				1 & 1 & 1 & 18 & 18 & 18 & 18 & 18 & 18 & 1 & 1 & 1 & 1 & 1 & 1 & 18 & 18 & 18 \\
			\end{pmatrix},
		\end{equation*}
		
		\vspace*{0.5cm}
		where the rows correspond to (the evaluator of) the basis $ \{1,x,x^3,x^6,x^7,x^9\} $ and the columns to the elements of $\mathbb{F}_{19}^*$ ordered as in Figure \ref{fig:float}.
		This means that to check the locality of each set one just needs to check the rank of the corresponding set of columns in the above matrix.
		For example, suppose we want to recover the third column, which corresponds to the symbol $11$. We can do that using only the first two columns, since the matrix
		\begin{equation*}
			\begin{pmatrix}
				1 & 1 & 1 \\
				1 & 7 & 11 \\
				1 & 1 & 1 \\
				1 & 1 & 1 \\
				1 & 7 & 11 \\
				1 & 1 & 1
			\end{pmatrix}
		\end{equation*}
		has rank equal to $2$.
		Similarly, we can recover any two lost symbols using either $3$ (if they belong to the same large orbit) or $4$ (otherwise, if they belong to distinct large orbits) other symbols.

		\section{Existential Results via Chebotarev Density Theorem}\label{sec:existresults}

In this section we explain how to apply Chebotarev Density Theorem to count the places $t_0\in \vF_q$ such that  $f(h)-t_0$ is totally split. A lower bound on this quantity determines directly a lower bound on the size of the hierarchy in our construction.
 This determines completely the range of parameters of our hierarchical codes, and in turn it shows that they always exist for $q$ large enough, without arithmetic restrictions on the localities and the size of the base field.

\subsection{Background on Galois theory}\label{sec:background_notions}
We begin by recalling a few preliminary definitions. Let $K$ and $M$ be fields. We will write $K[X]$ to denote the polynomial ring in the indeterminate $X$ over $K$. The field extension $K \subseteq M$ will be written as $M/K$, and its degree, that is, the dimension of $M$ as a $K$-vector space, as $[M : K]$. For $q$ a power of a prime, let $\mathbb{F}_q$ be the finite field with $q$ elements and let $\mathbb{F}_q^* = \mathbb{F}_q \setminus \{0\}$ be its (cyclic) multiplicative subgroup. Let $t$ be transcendental over $\vF_q$ and denote by $\vF_q(t)$ the rational function field in $t$ over $\vF_q$.

	We follow closely the notation and terminology in \cite{stichtenoth2009algebraic} throughout this section, and we provide the essential notions here. A finite-dimensional extension $K$ of $\vF_q(t)$ is called a (global) function field over $\vF_q$. A valuation ring of a function field $M/K$ is a ring $\cO$ such that $K \subsetneq \cO \subsetneq M$ and which contains at least one of $z$ or $z^{-1}$ for every $z \in M$. A place $P$ of $M/K$ is the unique maximal ideal of some valuation ring $\cO$ of $M/K$, and the degree of $P$ is defined to be $\deg P = [\cO/P : K]$. In particular, $P$ is called a rational place of $M/K$ if $[\cO/P : K] = 1$. There is a one-to-one correspondence between places of $M/K$ and valuation rings $\mathcal{O}$ of $M/K$, so we will write $\cO_P$ to denote the valuation ring whose maximal ideal is $P$. We will write $\mathbb{P}_M$ to denote the set of all places of $M/K$ and $\mathbb{P}_M^1 \subseteq \mathbb{P}_M$ to denote the set of rational places of $M/K$.
	Let $K \subseteq M$ be an extension of function fields. The automorphism group of $M/K$, that is, the group of all automorphisms of $M$ which fix $K$ element-wise, is denoted by $\Aut(M/K)$. When $|{\Aut}(M/K)| = [M : K]$, we say that the extension $M/K$ is Galois with Galois group $\Gal(M/K) = \Aut(M/K)$. For places $P \in \vP_K$ and $Q \in \vP_M$, we say that $Q$ lies above $P$ (and write $Q \mid P$) if $P \subseteq Q$. We denote the ramification index and relative degree of the extension of places $Q \mid P$ by $e(Q \mid P)$ and by $f(Q \mid P)$, respectively. We say that a polynomial $f \in \vF_q[X]$ is separable over $\vF_q$ if $f \not \in \vF_q[X^p]$, where $p = \textup{char} \, \vF_q$, and for such an $f$, the polynomial $f - t$ is seen to be a separable irreducible polynomial over $\vF_q(t)$. We will write $M_f$ to denote the splitting field of $f-t$ over $\vF_q(t)$. Equivalently, $M_f$ denotes the Galois closure of the extension $\vF_q(x):\vF_q(t)$, where $x$ is any root of $f(X) - t$ in the algebraic closure $\overline{\vF_q(t)}$ of $\vF_q(t)$. The field of constants of $M_f$ will be denoted by $k_f$, and we note that it is possible to have $k_f \supsetneq \vF_q$. Let $G_f$ be the monodromy group (sometimes called the arithmetic Galois group) of $f$, that is, the Galois group of the extension $M_f/\vF_q(t)$.

\subsection{The number of totally split places $t_0$ of $f(h)-t$}
We will appeal to the Chebotarev density theorem as in Proposition 3.1 of \cite{micheliIEEE} since this formulation is the most convenient for our purposes. We provide a full exposition in this section, but we briefly describe in the next paragraph the general procedure and ideas.

For polynomials $f,h \in \vF_q[X]$, consider the composition $f(h)$. By the lower bound in \cite[Proposition 3.1]{micheliIEEE} on the number $\ell$ of $t_0 \in \vF_q$ such that $f(h) - t_0$ splits into linear factors over $\vF_q$,  we have that for large enough $q$ it is guaranteed to have a large number of totally split places of degree $1$ of $\vF_q(x)/\vF_q(t)$ when $f(h)$ is chosen correctly. Now, we may assume that the field of constants $k_{f(h)}$ of $M_{f(h)}$ is trivial since otherwise there cannot be a totally split place of degree $1$. Since we want $\ell$ to be as large as possible, one quickly sees from the lower bound in \cite[Proposition 3.1]{micheliIEEE} that minimizing the size of the monodromy group $G_{f(h)}$ of $f(h)$ achieves this goal. Thus our construction always effectively results in an optimal code as long as the size of the alphabet verifies a certain lower bound.

		For the extension $M/\vF_q(t)$, let $G = \textup{Gal}(M/\mathbb{F}_q(t))$ be its arithmetic Galois group and let $N$ be its geometric Galois group. Since we are interested in the number $\ell$ of places $P \subseteq \mathbb{F}_q(t)$ of degree $1$ which are totally split in $M$, by Proposition 3.4 of \cite{micheliIEEE} we may assume that $M \cap \overline{\mathbb{F}}_q = \mathbb{F}_q$ is the field of constants of the extension $M/\mathbb{F}_q(t)$ since otherwise $\ell = 0$. Hence $G = N$.
		
		\begin{lemma}\label{lemma:tot_split}
			Let $f,h \in \mathbb{F}_q[X]$ be nonzero polynomials having positive degrees. Define $G_f = \textup{Gal}(f(X)-t : \mathbb{F}_q(t))$ and similarly for $G_h$. Then the number of $t_0 \in \mathbb{F}_q$ such that $f(h(X)) - t_0$ splits completely into distinct (linear) factors over $\mathbb{F}_q$ is at least $\displaystyle \frac{1}{|G_h|^{\deg f} |G_f|} q + O(\sqrt{q})$, where the implied constant can be chosen explicitly and is independent of $q$.
		\end{lemma}
		
		\begin{IEEEproof}
			Denoting the number of $t_0 \in \mathbb{F}_q$ we are considering by $|T^1_{split}(f \circ h)|$, from Proposition 3.1(ii) of \cite{micheliIEEE} we immediately have 
\begin{equation}\label{eq:bound}		
|T^1_{split}(f \circ h)| \geq \displaystyle \frac{q+1-2g\sqrt{q}}{|G|} - \frac{\# \textup{Ram}^1(M : \mathbb{F}_q(t))}{2}.
\end{equation} 
We proceed by proving an upper bound on the size of $G$, which in turn gives the wanted lower bound for $|T^1_{split}(f \circ h)|$
			
			Let $\mathcal{T}$ be the rooted tree of height $2$ with $\deg f$ branches and $\deg h$ roots adjacent to each branch. One can easily see that $G \subseteq \textup{Aut}(\mathcal{T})$, so because $\textup{Aut}(\mathcal{T})$ is isomorphic to the wreath product $(\underbrace{G_h \times \cdots \times G_h}_{\deg f}) \rtimes G_f$, we have $|G| \leq |G_h|^{\deg f} |G_f|$.
			
			Combining \eqref{eq:bound} with the bound on $|G|$, we obtain $$\ell \geq \frac{q+1 - 2g\sqrt{q}}{|G_h|^{\deg f} |G_f|} - \frac{\textup{Ram}^1(M:\mathbb{F}_q(t))}{2}.$$
		\end{IEEEproof}
		
		Note that the bound given in the previous lemma can be written more explicitly as $$\ell \geq \frac{(q+1) - 2g\sqrt{q}}{|G_h|^{\deg f}|G_f|} - (\deg f)(\deg h)/2.$$
		
\begin{proposition}\label{prop:existprop}
Let $f,h\in \vF_q[x]$ be polynomials such that $f(h)-t$
 has Galois group $G$ and the splitting field $M$ of $f(h)-t$
 has constant field equal to $\vF_q$. 
Then there exists an optimal H-LRC, with parameters 
$[\deg(f(h))\ell,k,d,\deg(h)-1,\deg(f(h))-\lambda,\lambda ]$ for any $\lambda<\deg(h)$ and \[\ell\geq\frac{ q}{|G_h|^{\deg f} |G_f|} +O(\sqrt q),\] where the implied constant can be made explicit, and $G_h$ (resp $G_f$) is the Galois group of $f-t$ (resp. $h-t$), and $k$ is as in Lemma \ref{lm:k}.
\end{proposition}
\begin{remark}
Notice that the condition of having trivial constant field extension is automatic once there is a single totally split place, and it is the generic situation if the polynomials are chosen at random.
\end{remark}
\begin{IEEEproof}
Since $M$ has trivial constant field $\vF_q$, Lemma  \ref{lemma:tot_split} guarantees that there exist at least 
\[\ell\geq \frac{1}{|G_h|^{\deg f} |G_f|} q + O(\sqrt{q})\]
totally split places, i.e. elements $t_0$ of $\vF_q$ such that $f(h)-t_0$ is totally split. Now Lemma \ref{lm:f(h(X))_split} guarantees that the evaluation set $T$ consisting of the preimages of the $t_0$'s forms a nest for the pair $(f,h)$ (see Remark \ref{rem:subnest}).
Now construct the code by evaluating the polynomials in \eqref{eq:message} at the subset $\mathcal A$ of preimages of $T$ via $f(h)$, i.e. $\mathcal A=(f\circ h)^{-1}(T)$, which has size $\deg(f(h))\ell$. The hierarchy is now given by the nest structure in the sense of Remark \ref{rem:subnest}, and the parameters obtained from Section \ref{sec:codereal}. 
\end{IEEEproof}
		
\section{Practical choice of parameters to construct optimal HLRC}\label{sec:practical}
		The construction we presented in the previous sections allows us to exhibit some interesting examples of HLRCs. To begin with, we consider the case $\mathbb{F}_{64}$. Choosing $f$ and $h$ such that $\deg f= \deg h=3$ and $\ell=7$, our construction gives rise to a $(63,k,d,2,5,2)$-HLRC, where the values of $k$ and $d$ depend on the choice of $s$ in Construction \ref{con:LRC}. In fact, the first locality $b$ equals $\deg h -1$, whereas the second locality ($a=5$) can be computed by following the passages of Section \ref{sec:locality}.
		This means that we are able to recover either $ 1 $ (resp. $ 2 $) lost node(s) by looking either at $2$ (resp. $5$) other nodes. 
		We point out that the Tamo-Barg construction for availability over the field of size $64$, under the same first locality assumption ($ b=2 $), forces to have length $21$ (with locality sets of size $3$ and $7$), whereas ours permits to have lenght $63$, leading to a much better minimum distance and a larger number of servers allowed. More precisely, the Tamo-Barg construction requires the use of two orthogonal partitions, and this can be achieved by using $21$ symbols corresponding to the action of $x^3$ and of $x^7$ on $\mathbb{F}_{64}\setminus \{0\}$. Note further that their construction has a larger second locality: $7$, against our better parameter $a=5$.

\section{Acknowledgements}
This research is supported by the National Science Foundation under Grant No. 2127742.

\bibliographystyle{abbrv}
\bibliography{biblio5.bib}

\begin{IEEEbiography}{Austin Dukes}
is a graduate student at the University of South Florida working in Applied Algebra and Coding Theory.
\end{IEEEbiography}
\begin{IEEEbiography}{Giacomo Micheli}
graduated at the University of Rome ``La Sapienza'' in July 2012. He completed his Ph.D. with distinction at the Zurich Graduate School in Mathematics in October 2015 under the supervision of Prof. Joachim Rosenthal. He is currently a tenure-track assistant professor at the University of South Florida and Co-Director of the Center for Cryptographic Research at USF.
\end{IEEEbiography}
\begin{IEEEbiography}{Vincenzo Pallozzi Lavorante}
graduated at the University of Perugia in October 2018. He completed his Ph.D. with distinction at the University of Modena and Reggio Emilia in February 2022 under the supervision of Prof. Massimo Giulietti. He is currently a Post-Doc researcher at the University of South Florida.
\end{IEEEbiography}
	\end{document}